\documentclass[prl,twocolumn,amssymb,showpacs]{revtex4}
\usepackage[T1]{fontenc}
\usepackage[latin1]{inputenc}
\usepackage{graphics}



\begin{document}

\title{Entropy Analysis of Stochastic Processes at Finite Resolution}
\author{D.M. Tavares}
\author{L.S. Lucena}
\affiliation{International Center for Complex Systems}
\address{Departamento de Física Teórica e Experimental - UFRN \\
Natal-RN 59078-970 Brazil}

\begin{abstract}
The time evolution of complex systems usually can be described
through stochastic processes. These processes are measured at
finite resolution, what necessarily reduces them to finite
sequences of real numbers. In order to relate these data sets to
realizations of the original stochastic processes (to any
functions, indeed) it is obligatory to choose an interpolation
space (for example, the space of band-limited functions). Clearly,
this choice is crucial if the intent is to approximate optimally
the original processes inside the interval of measurement. Here we
argue that discrete wavelets are suitable to this end. The wavelet
approximations of stochastic processes allow us to define an
entropy measure for the order-disorder balance of evolution
regimes of complex systems, where order is understood as
confinement of energy in simple local modes. We calculate exact
results for the fractional Brownian motion (fBm), with application
to Kolmogorov K41 theory for fully developed turbulence.
\end{abstract}


\maketitle

Many physical systems investigated at present have a complex
evolution in time. Frequently, the major information we can obtain
on their dynamics comes from time series of noisy
appearance~\cite{consolini,politzer,santi,sreenivasan,gafarov}.
These series are samples at finite resolution of subjacent
stochastic processes, whose properties are better investigated
through a multi-resolution approach, because the realizations of
the mentioned stochastic processes are singular everywhere
functions. This is, for example, the case of \(1/f^{\alpha}\)
noises from self-organized
systems~\cite{bak,durin,urbach,cote,carreras}. Thus, singularities
are common, and should be interpreted as details that influence
function variation at all scales~\cite{muzy,mallat}. In this paper
we propose a method for entropy analysis of arbitrary complex time
series that takes advantage of this theoretical standpoint.
Moreover, we account for the fact that measurements are made at
finite resolution, considering the consequences of sampling, what
is not fully accomplished in previous approaches. In general,
other formulations are strongly influenced by
information-theoretical arguments, applied to the analysis of
chaotic behavior~\cite{bandt}. We see difficulties in two main
aspects. First, the notion of complexity that is employed in these
formulations is based on entropy rates, like the Kolmogorov-Sinai
entropy, that measure the degree in which information on initial
conditions is lost when the systems evolve. This gives a scale of
complexity ranging from zero (non-chaotic deterministic case) to
infinite (stochastic case). In this scale, finite values quantify
the complexity of deterministic chaos. As a consequence, the
problem of defining a proper complexity estimator for stochastic
processes is substituted by the statistical investigation of
chaotic deterministic dynamical systems~\cite{eckmann}. Second,
these formulations do not consider the relationship that scaling
has with disorder~\cite{madalena}. Our approach is tailored to
face these difficulties. Here, we begin by assuming that the
stochastic processes are supported on the real axis, and measured
at a discrete set of points with a sampling interval \( \tau \),
during a time \(T\). The resolution of the measurement is \(
N=T/\tau \), and for convenience we make \( T=1 \), and \( \tau
=1/N=2^{-J} \) (we will work in these units). Such measurement
results in a loss of information, which depends on two factors:
The resolution, and the interpolation space in which the
stochastic processes will be projected~\cite{unser}. The idea is
that the mere sampled values say almost nothing about a function.
Much more information is conveyed through the hypothesis on how
the function varies between the sampled values. This regularity
hypothesis, usually implicit when we draw smooth curves between
data points, is an essential element of the theory. Without this
hypothesis, no information found on the discrete and finite data
sets can be attributed to the subjacent model, that is assumed to
hold on the continuous support. Now, in the time-scale
(time-frequency) domain, it makes sense to search a representation
\emph{i}) that is minimally affected by the created end
singularities, \emph{ii}) that provides an interpolation based on
multi-scale approximation of the actual singularities, and
\emph{iii}) that deals with the resolution \(N\) as a direct
experimental parameter. The first and second requirements are the
most crucial, and establish the way in which the energy
(\(L^2\)-norm squared) is assumed to be distributed inside the
interval of measurement, so that this distribution corresponds to
the singularity structure of the process, seen at a finite
resolution. As a supplementary condition, the best is that the
algorithmic complexity grows only linearly with the length of the
time series. These requirements are met if we project the
stochastic processes in a discrete wavelet space.

Wavelets are associated with time-scale representations. Let \(
\left| F\right\rangle  \) be a vector in a Hilbert space. There
are bases \( \left\{ \left| \psi _{jn}\right\rangle \right\}  \)
such that \( \left| F\right\rangle \) can be expanded as

\begin{equation}
\label{eq:waveletexpansion1} \left| F\right\rangle =\sum
^{j=\infty }_{j=-\infty }\sum ^{n=\infty }_{n=-\infty
}\left\langle \psi _{jn}|F\right\rangle \left| \psi
_{jn}\right\rangle ,
\end{equation}
the form of the basis functions being \( \psi _{jn}\left( t\right)
=\frac{1}{\sqrt{2^{j}}}\psi \left( \frac{t-2^{j}n}{2^{j}}\right) .
\) These functions are called discrete wavelets, and are square
integrable with zero mean, and may have \(p\) vanishing moments.
The discrete wavelet transform of \( F \) to this basis is written
\( D_{jn}F\equiv\left\langle \psi _{jn}|F\right\rangle ,\) and
gives information on the behavior of the function \( F \) at scale
\( 2^{j} \) and time \( 2^{j}n \)~\cite{mallat}.

\begin{figure}
{\par\centering \resizebox*{6cm}{6cm}{\includegraphics{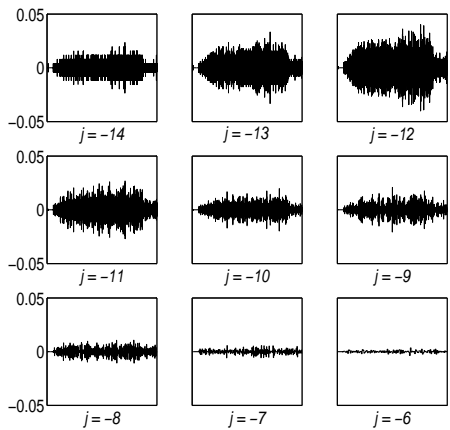}}
\par}
\caption{The Haar wavelet decomposition of a stochastic process
sampled at the finite resolution \(N=2^{15}\). The details are
presented from scale the \(2^{-14}\) to the scale \(2^{-6}\). Note
how the energy accumulates at the small scale details. This causes
the high entropy \(S=12.52\) as compared to the maximum
\(S_{\rm{max}}=15\).\label{fig:coefstable}}
\end{figure}

Confining the function in expression~(\ref{eq:waveletexpansion1})
to a finite observation window, and measuring at finite
resolution, we obtain the approximation
\begin{equation}
\label{eq:waveletexpansion2} \left| f_J\right\rangle =\left\langle
\phi |f\right\rangle \left| \phi \right\rangle +\sum
^{j=0}_{j=-J+1}\sum ^{n=2^{-j}-1}_{n=0}\left\langle \psi
_{jn}|f\right\rangle \left| \psi _{jn}\right\rangle ,
\end{equation}
 where \( f \) is the restriction of \( F \) to \( [0,1] \).
The function \( \phi  \) is called the scaling function, having
the properties of a smoothing kernel~\cite{mallat}. We are
interested in the fluctuations, so, without any loss, we employ in
our analysis the transform \(f\rightarrow f-\left\langle
f\right\rangle\), getting from \(f\) a version with null
time-average. At this point we are ready to express an energy and
an entropy for the time series, that are also resolution dependent
approximations of the energy density and entropy density for the
ideal stochastic process. The energy of \( f_J \) is defined by
the ensemble average of the \(L^2\)-norm squared,
\begin{equation}
\label{eq:energydensity} E\left(J\right)
\equiv\overline{\left\langle f_{J}|f_{J}\right\rangle },
\end{equation}
that is directly determined with the simple formula \(
E\left(J\right) =\overline{\sum ^{i=N}_{i=1}f^{2}_{i}}, \) where
\( f_{i} \) are the sampled values of \( f \). The entropy has the
definition
\begin{equation}
\label{eq:entropydensity} S\left(J\right) \equiv -\sum
^{j=0}_{j=-J+1}\sum ^{2^{-j}-1}_{n=0}P_{jn}\log
_{2}P_{jn}-P_{\phi}\log_2P_{\phi},
\end{equation}
with \(P_{jn}\equiv \overline{\ |D_{jn}F|^2}/E\left(J\right)\),
and \(P_{\phi}\equiv \overline{\ |\left\langle
\phi|f\right\rangle|^2}/E\left(J\right)\). This expression has
similarities with the basis entropy cost, using the language of
reference~\cite{mallat}. In the limit of a single pulse localized
around \(t\) and with duration \(\delta t\sim k\tau\), with \(k\)
a natural number, only very few wavelet coefficients around the
scale \(k\tau\) and localized near \(t\) will be appreciably
different from zero. This characterizes low entropy processes. On
the other hand, the white noises~(WN) in which the energy is, in
average, equally distributed on all wavelets (the same power
dissipation at all times and scales) have the maximum entropy
\(S_{\rm{max}}=J\). This entropy functional is useful to quantify
the balance between order and disorder (complexity) for different
regimes of evolution of complex systems. Actually, we expect that
the level of disorder in a process mainly depends on the energy
distributed into the details. Consider, for example, the
interaction between two surfaces that slip one over the other.
When there is no friction we expect no sound. But if we introduce
roughness, the energy will be dissipated in microscopic
collisions, originating a sound that is classified as a noise, due
to the excess of details. Indeed, the intervals between such
collisions are so small that no measurement will capture them. The
process will continue a noise for \lq high\rq\ resolution
instruments, like it is for our \lq low\rq\ resolution ears. In
FIG.~\ref{fig:coefstable} we see the wavelet decomposition of the
sound caused by a piece of plastic pressed and pushed on a rough
table. The calculated entropy was \(S=12.52\) for a maximum of
\(S_{\rm{max}}=15\). The random small scale details have
amplitudes comparable to that of the function during the whole
observation, constituting important components. The great amount
of energy distributed on the multitude of details causes this high
entropy. This is why we suggest that the proposed entropy is
physically appropriate.

A paradigmatic model for the self-affine processes is the
fractional Brownian motion (fBm), introduced by Mandelbrot and Van
Ness in the realm of stochastic processes with \( 1/f^{\alpha } \)
spectra~\cite{mandelbrot}. The fBm is a mono-parametric stochastic
process \( B_{H} \) with null ensemble average, Gaussian
increments, and whose correlation function has the general form:
\begin{equation}
\label{eq:correlation} \overline{B_{H}\left( t\right) B_{H}\left(
s\right) }=\frac{\sigma ^{2}_{H}}{2}\left( \left| t\right|
^{2H}+\left| s\right| ^{2H}-\left| t-s\right| ^{2H}\right) ,
\end{equation}
with \(\sigma ^{2}_{H}\equiv \overline{B^{2}_{H}\left( 1\right)
}\). The parameter of the process is the Hurst exponent \( H
\in(0,1)\), which controls the nature of the correlations. The
variance of \( B_{H} \) depends on time according to
\(\overline{B^{2}_{H}}=\sigma ^{2}_{H}\left| t\right| ^{2H},\)
that reduces to the Brownian motion form when \( H=1/2 \),
representing the uncorrelated case. In this process, for \( H<1/2
\) the increments present infinite range anti-correlations, while
for \( H>1/2 \) there are infinite range correlations. The Fourier
spectrum of fBm follows the power law
\begin{equation}
\label{eq:spectrum}  S\left( f \right) \sim \frac{1}{f ^{2H+1}}.
\end{equation}

\begin{figure}
{\par\centering \resizebox*{6cm}{6cm}{\includegraphics{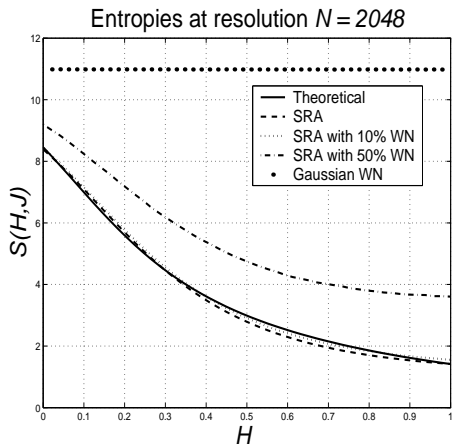}}
\par}
\caption{The solid line is the theoretical fBm entropy at
resolution \protect\( N=2048\protect \). We see that the SRA
algorithm gives a very good approximation to the theoretical
prediction. Two lines represent the effect of Gaussian white noise
superposition. The (visually) straight dotted line at the upper
part gives the entropy of simulated pure Gaussian white
noise.\label{fig:SversusH}}
\end{figure}

In the Haar wavelet basis one finds exact and simple expressions
for the resolution dependent energy and entropy densities of fBm.
The Haar wavelets are periods of square waves written as
\begin{equation}
\label{eq:haarwavelets} h_{jn}\left( t\right) =\left\{
\begin{array}{cc}
\frac{1}{\sqrt{2^{j}}}, & 2^{j}n\leq t<2^{j}\left( n+1/2\right) \\
-\frac{1}{\sqrt{2^{j}}}, & 2^{j}\left( n+1/2\right) \leq
t<2^{j}\left( n+1\right)
\end{array}\right. .
\end{equation}
 For Haar wavelets one proves, using the expression~(\ref{eq:correlation}), that
\begin{equation}
\label{eq:expectedvalueofWTsquared} \overline{\left| \Delta
_{jn}b_{H,J}\right| ^{2}}=\frac{\sigma ^{2}_{H}\left(
1-2^{-2H}\right) }{\left( 2H+1\right) \left( 2H+2\right)
}2^{\left( 2H+1\right) j},
\end{equation}
 where \( \Delta _{jn}f\equiv \left\langle h_{jn}|f\right\rangle  \), and \(b_H\) indicates restriction of
\(B_H\) to \([0,1]\). This expression gives the scale dependent
power spectrum in the Haar basis, from which one can calculate the
energy density of \( B_{H} \)
\begin{equation}
\label{eq:energydensity} E\left( H,J\right)
=\overline{\left\langle b_{H,J}|b_{H,J}\right\rangle
}=\frac{\sigma ^{2}_{H}\left( 1-2^{-2HJ }\right) }{\left(
2H+1\right) \left( 2H+2\right) },
\end{equation}
depending on \(\sigma_H\), that must be determined experimentally
. The entropy density has the final form
\begin{widetext}
\begin{equation}
\label{eq:shannonresult} S\left(H,J\right )=2^{-2H}\left(
2H+1\right) \frac{1-2^{-2H(J-1)}J +2^{-2HJ}(J-1)}{\left(
1-2^{-2H}\right) \left[ 1-2^{-2HJ }\right] }+\log _{2}C\left(
H,J\right) ,
\end{equation}
\end{widetext}
 where \(C\left(H,J\right)=\left[1-2^{-2HJ}\right]/
 \left(1-2^{-2H}\right)\).

 This entropy density is universal and exact for the
fBm processes sampled without rounding error to a given finite
resolution. It is very useful since for a great number of
practical applications there are just low resolution data, what
makes mandatory that the theoretical predictions take resolution
into account explicitly. In the limit \(J\rightarrow\infty\) the
entropy diverges at \(H=0\) as \(S(H,\infty)\sim 1/2H\ln 2\). For
other values of \(H\) the entropy is finite at infinite
resolution. At finite resolution \(S(0,J)=(J-1)/2+\log_2(J)\).

The FIG.~\ref{fig:SversusH} shows a plot of fBm theoretical
entropy for the resolution \( N=2048 \). The theoretical result is
compared with \( S\left( H,J\right)  \) calculated from samples
simulated with the Successive Random Additions (SRA)
algorithm~\cite{peitgen}. It shows also how Gaussian white
noise~(WN) affects the entropy, clarifying what is to be expected
from noisy data. For comparison, the dots at the upper part of the
figure show the entropy for the simulated pure Gaussian white
noise, whose average value is \(S=10.98\sim 11\). In the dotted
line we see the effect of a superposed noise with 10\% of the
amplitude of the fBm. The dash-dotted line shows this effect when
the percentage grows to 50\%. The entropies for all simulated
processes were calculated from 100 samples. The drastic diminution
of entropy when \(H\) increases happens because the details
(roughness) become weaker. There is very good agreement between
theoretical prediction and simulation.

In conclusion, we propose that the entropy
measure~(\ref{eq:entropydensity}) is suitable to estimate the
complexity of time series by revealing the degree in which
\emph{detail modes} are excited in a process. These detail modes
are easily represented as localized oscillations or wavelets,
which in turn are very effective mathematical instruments for
time-scale analysis. Considering such phenomena as fluid or scalar
turbulence~\cite{zsshe,shraiman}, that are characterized by
fluctuating cascades from high inertial scales to low dissipative
scales, or plasma turbulence~\cite{krommes}, in which some scales
are most important for the energy transfer, the proposed entropy
seems a natural measure of the intrinsic order-disorder balance.
In such phenomena, many structures (vortices, convective cells)
appear at intermediate scales, that should be detected as
excitations of intermediate scale modes, lowering the entropy, as
compared with noises that are completely determined by microscopic
dissipation (whose characteristics are indicated in
FIG.~\ref{fig:coefstable}). For example, if a theory with the same
correlations of Kolmogorov K41 theory~\cite{kolmogorov41} is
valid, implying \(H=1/3\), the time series of velocity increments
would lead to the entropy \(S(1/3,\infty)=4.27\dots\), a number
significantly small, compared to the case in
FIG.~\ref{fig:coefstable}. What we learn from the fBm analysis
presented here suggests that the measurement of the entropy is a
strong method of characterization. Furthermore, it is interesting
to obtain experimental curves of entropy for stochastic processes
depending on parameters, or in situations where the entropy may
change with time. This method has given results (which will be
published soon) in three situations: In the analysis of the sound
of lungs, in the analysis of heart beat series, and in the
analysis of global positioning system errors caused by equatorial
spread F in ionospheric plasma. We observe that such entropy
measurements could give important information on the
non-equilibrium approach to self-organized regimes~\cite{corral},
as well as, on self-organized regimes made unstable, for example,
by diffusion~\cite{newman}.

We acknowledge the CAPES, CNPq, FINEP and CTPETRO for financial
support.

\end{document}